\documentclass[11pt, runningheads,a4paper]{llncs}

\usepackage{graphicx}
\usepackage{geometry}
\usepackage{amsmath}
\usepackage[utf8]{inputenc}
\usepackage[default]{lato}
\usepackage[T1]{fontenc}

\usepackage{etoolbox}
\patchcmd{\thebibliography}{\section*{\refname}}{}{}{}
\let\OLDthebibliography\thebibliography
\renewcommand\thebibliography[1]{
  \OLDthebibliography{#1}
  \setlength{\parskip}{0pt}
  \setlength{\itemsep}{0pt plus 0.3ex}
}
\usepackage{xcolor}
\usepackage{amssymb}
\usepackage{graphicx}

\usepackage{natbib}

\usepackage{url}

\usepackage{parskip}
\usepackage{datetime}
\begin{document}

\mainmatter  

\title{Unravelling the forces underlying urban industrial agglomeration}

\author{Neave O'Clery\and Samuel Heroy\and Fran\c{c}ois Hulot\and Mariano Beguerisse-D\'iaz}

\institute{Mathematical Institute, University of Oxford}

\maketitle

\begin{center}
   \footnotesize{\today}
\end{center}

\vspace{0.3cm}
\noindent\rule{\linewidth}{0.35pt}
\vspace{0.3cm}

A B S T R A C T

\vspace{0.3cm}
\noindent\rule{\linewidth}{0.35pt}
\vspace{0.3cm} 

As early as the 1920's Marshall suggested that firms co-locate in cities to reduce the costs of moving goods, people, and ideas. These 'forces of agglomeration' have given rise, for example, to the high tech clusters of San Francisco and Boston, and the automobile cluster in Detroit. Yet, despite its importance for city planners and industrial policy-makers, until recently there has been little success in estimating the \textit{relative importance} of each Marshallian channel to the location decisions of firms. 

Here we explore a burgeoning literature that aims to exploit the co-location patterns of industries in cities in order to disentangle the relationship between industry co-agglomeration and customer/supplier, labour and idea sharing. Building on previous approaches that focus on across- and between-industry estimates, we propose a network-based method to estimate the relative importance of each Marshallian channel at a meso scale. Specifically, we use a community detection technique to construct a hierarchical decomposition of the full set of industries into clusters based on co-agglomeration patterns, and show that these industry clusters exhibit distinct patterns in terms of their relative reliance on individual Marshallian channels.

\vspace{0.3cm}
\noindent\rule{\linewidth}{0.35pt}
\vspace{0.3cm}

\section{Introduction}

Globally, a consensus has emerged that emphasises the role of cities, rather than countries, as key engines of economic growth \citep{jacobs1961death,glaeser2011triumph}. Yet, what are the drivers of urban success in the face of high costs arising from factors such as congestion, density, crime, and pollution? 

Leading scholars have argued that it is the diversity of cities, and in particular the way cities facilitate and foster a diverse ecology of social interactions, that gives rise to new activities, opportunities and innovations \citep{jacobs2016economy, bettencourt2007growth}. This view aligns with a growing literature that emphasises the role that larger cities have in better facilitating matching between employers and employees, knowledge spillovers between firms and innovation opportunities \citep{Friedrichs1993theory, DurantonPuga2001NurseryCities, DurantonPuga2004micro, RosenthalStrange2006}. 

Yet, for all their diversity at an individual or worker level, cities tend to specialize at an industry level \citep{beaudry2009s}. In other words, industries are geographically concentrated: examples range from the tech cluster of Silicon Valley to the automobile cluster of "motor city" (Detroit). This phenomenon of clustering of similar firms is thought to arise mainly via three channels of cost-sharing: transport costs associated with customers and suppliers, labour costs, and learning or knowledge costs \citep{marshall1920economics,glaeser1992growth,henderson1995industrial}. 

However, until recently there has been little success in estimating the \textit{relative importance} of each of these channels to the location decisions of firms. This chapter will explore and extend a burgeoning literature \citep{EllisonGlaeserKerr2010,diodato2018industries} that aims to exploit the co-location patterns of industries in cities in order to unravel the relationship between industry co-agglomeration and customer/supplier, labour and idea cost-sharing. 

%%%%%%%%%%%%%%%%%%%%%%%%%%%%%%%%%%%%%%%%%%%%%%%%%%%%%%%%%%%%%%%%%%%%%%%%%%%%%%%%%%%%%%%%%%%
\subsection{Urban agglomeration patterns}

The benefits of firm agglomeration have long been proposed. In 1920, Marshall argued that firms locate in close proximity in order to reduce costs: firms can save shipping costs by locating near suppliers or customers, searching and matching costs via labour market pooling, and benefit from knowledge spillovers \citep{marshall1920economics}. For example, chemical and pharmaceutical companies might rely on access to a similar labour pool, whereas car manufacturers might be more sensitive to transport costs associated with suppliers of parts. Industries at the forefront of innovation such as biotechnology firms might benefit from a combination of skilled labour and knowledge spillovers from locating close to other similar firms. In a well-known instance of this type of location decision, Amazon was set up in Seattle to benefit from the local talent pool (i.e., Microsoft and other tech firms).

Other benefits of geographical clustering of firms identified in literature are competition, which drives productivity \citep{porter2011competitive}, and local demand effects \citep{Fujiteetal2001}. 

While it has been long understood that it is these 'externalities' that drive urban industry agglomeration patterns, measuring their impact, and modelling their behaviour and dynamics, remains a challenge. In particular, until quite recently there has been little success in estimating the \textit{relative importance} of each Marshallian channel to firms' decisions - not least because each channel predicts an identical agglomeration pattern \citep{DurantonPuga2004micro}. 

In response to this challenge, a pioneering study by \citet{EllisonGlaeserKerr2010} (henceforth EGK) proposed to exploit the co-location patterns of industries in cities in order to estimate the relationship between industry co-agglomeration and each of the three Marshallian channels: customer/supplier, labour and idea sharing. In other words, they were interested in whether pairwise patterns of industry co-agglomeration were strongly correlated with, for example, labour pooling (which can be constructed from an estimate of skill-overlap between industries, see next section). If co-agglomeration patterns are most correlated with labour pooling (compared to the other channels), then, EGK propose, the labour channel is the most important reason for firm location decisions \textit{across all industries.} Hence, EGK argue that comparing co-agglomeration patterns to each of the three channels enables us to estimate the relative importance of each channel (across all industries).

In practice, this means constructing a pairwise industry-industry measure of co-location (e.g., the propensity of firms from two industries to be located in the same cities), and corresponding pairwise estimates for each industry-industry 'proximity' or similarity for each of the three Marshallian channels. There are a variety of ways to measure these similarities as discussed below.

%%%%%%%%%%%%%%%%%%%%%%%%%%%%%%%%%%%%%%%%%%%%%%%%%%%%%%%%%%%%%%%%%%%%%%%%%%%%%%%%%%%%%%%%%%%
\subsection{Industry similarity and networks\label{sect_sim}}

For each of the studies we focus on in this chapter, data on industries and employment from the USA (predominantly from 2002) is used. Technical definitions and information on data sources are provided in Section 2. Following EGK, our task is to develop pairwise industry-industry measures of co-agglomeration, labour pooling, custom-supplier and knowledge sharing. 
 
Geographic co-agglomeration of industries is typically estimated using data on the distribution of employment by industry and city \citep{EllisonGlaeserKerr2010,hausmann2014}. 
 
Two industries are 'proximate' in terms of the first channel, labour pooling, if they have similar skill requirements. Hence, one method to estimate shared labour needs is to compute a measure of occupational similarity (e.g., using data on industry employment by occupation) \citep{farjoun1994beyond, chang1996evolutionary}. Alternatively, labour flows (job switches) between industry pairs is an excellent proxy for shared skills \citep{Neffke2013SkillRelatedness}. Unfortunately, sufficiently dis-aggregated data on industry-industry labour flows is unavailable for the USA at this point.
 
Two industries are similar in terms of the second channel if they share customers and suppliers. In order to capture buyer-seller relationships between industry pairs \citep{fan2000measurement}, we can make use of Input-Output matrices, built by national statistics offices worldwide, which capture monetary flows between industries (normally with great precision). 

Finally, a pair of industries is similar in terms of the third channel if they commonly exchange ideas and know-how (often via some sort of R\&D collaboration). This type of knowledge sharing between industries is typically captured using patent data - more specifically, the extent to which technologies associated with industry $i$ cite technologies associated with industry $j$, and vice versa. However, only a select few manufacturing industries are active in patenting \citep{diodato2018industries} and hence this third channel is omitted from much of the analysis below. 

It is instructive to think about each type of industry proximity (co-agglomeration, labour pooling, customer-supplier and knowledge sharing) in terms of a network, where nodes (industries) are connected via edges with weight equal to the corresponding proximity. In section \ref{section_div} we will see that industry networks of this type are frequently used to model diversification paths, whereby places (e.g., cities) branch into new economic activities that are similar to existing strengths \citep{frenken2007, hidalgo2007}. Later, in Section 4, we will propose a network-based method to estimate the relative importance of each Marshallian channel at an industry-cluster scale.

Network models are increasingly used to understand the role that interconnected structures play in economic and innovation-related processes, including research clusters \citep{catini2015}, the inter-industry propagation of supply and demand shocks \citep{acemoglu2015systemic}, and banking crises \citep{battiston2012debtrank}.

%%%%%%%%%%%%%%%%%%%%%%%%%%%%%%%%%%%%%%%%%%%%%%%%%%%%%%%%%%%%%%%%%%%%
\subsection{Relative importance of Marshallian channels}

Using the pairwise proximity matrices introduced in the previous section, EGK estimated the contribution of labour, customer-supplier linkages, and knowledge spillovers to co-location using 1987 manufacturing data from the US. The technical details of this estimation are provided in Section \ref{sect_EGK}. The authors find that each channel is a significant factor in industry co-location patterns, with a particular emphasis on customer-supplier linkages. 

\citet*{diodato2018industries} (henceforth DNO) reproduced the EGK study using data for US manufacturing industries from 2002, confirming a strong role for labour and input output linkages in co-agglomeration patterns. Evidence for the importance of knowledge spillovers was less apparent. These studies however focus on manufacturing industries and neglect services, an increasingly dominant component of advanced urban economies. Services, much like high and low skilled manufacturing, may favour labour linkages (services are often labour intensive) or customer-supplier linkages (many services are non-traded). Extending the EGK study, DNO found a stronger role for labour linkages when a larger set of industries including services were considered. 

Investigating further the strengthening role of labour linkages for co-agglomeration patterns, using data from 1910 to 2010 DNO showed that the importance of customer-supplier linkages has steadily decreased over the past century while labour linkages have strengthened. This result corresponds to a well-documented  shift from value chain based manufacturing to a specialised service driven economy. 

The framework proposed by EGK assumes a homogeneous contribution of each Marshallian channel to co-agglomeration patterns \textit{for all industries}. Intuitively, however, it is  highly likely that individual industries will rely on each Marshallian channel in different combinations. For example, as discussed above, one might expect that high tech manufacturing relies on the availability of skilled labour whereas heavy manufacturing prioritizes proximity to suppliers and customers. DNO adapt the framework proposed by EGK, and estimate the relationship between co-agglomeration patterns and each of the three Marshallian channels \textit{for individual industries}. Their results highlight remarkable heterogeneity across industries, with arts, media and scientific industries showing a strong preference for labour linkages, while farming and manufacturing co-locate according to customer and supplier linkages. DNO is one of a number of studies to investigate heterogeneity in agglomeration forces across industries \citep{faggio2014heterogeneous, howard2015measuring, behrens2016agglomeration}.

%%%%%%%%%%%%%%%%%%%%%%%%%%%%%%%%%%%%%%%%%%%%%%%%%%%%%%%%%%%%%%%%%%%%%%%%%%%%%%%%%%%%%%%%%%%
\subsection{Agglomeration drivers for industry clusters}

We can view previous approaches to measuring the relative importance of Marshallian channels as `local' approaches, which inherently do not take into account network structure. Specifically, previous approaches have sought to relate the whole network (e.g., all of the edges) or individual nodes (e.g., all of the edges connected to a single node) to patterns corresponding to each of the Marshallian channels. Each of these methods have neglected to consider intermediate scales which emerge naturally from the structure of the co-agglomeration network.

Intuitively, one might expect that clusters of industries co-located across many cities may exhibit a similar dependence on each of the Marshallian channels. We can identify such clusters via an analysis of connectivity patterns in the co-agglomeration network (henceforth EG network). Then, building on the framework established by EGK and DNO, we will show that industry clusters exhibit heterogeneous dependence on each of the Marshallian channels.  

In order to identify industry clusters, we will focus on uncovering densely connected groups of nodes (known as communities) in the co-agglomeration network. This approach is connected to work on detecting and defining industrial clusters for the US \citep{porter1998clusters,delgado2015clusters} and Ireland \citep{ocleryIreland}. More broadly, community detection has been used extensively to study the structure and dynamics of biological and social networks \citep{girvan2002community}.

Although there exist a wide range of approaches to community detection (see \citet{fortunato2010community} for a review), we will employ a method called \textit{Markov Stability} based on diffusion dynamics \citep{delvenne2010stability}. Intuitively, we let a random walker wander on the network, jumping from node to node. If the random walker remains in the same group of nodes over a long period of time, there is high connectivity between the group of nodes and a community has been discovered.  

While most well-known methods for community detection seek to find a single node partition under a particular optimisation strategy (e.g., modularity from \citet{blondel2008fast}), it is more natural to think about a range of partitions on different scales (from many small node clusters to few larger clusters). Within the prism of the Stability approach, this information can be extracted by analysing the patterns of random walkers on a network at different timescales \citep{delvenne2010stability}. Longer time-scales correspond to larger node aggregations, and fewer communities. 

We apply the Stability method to the co-agglomeration network, extracting a range of network partitions at different scales. Although not strictly nested, these partitions form a hierarchical decomposition of the co-agglomeration network into industry clusters of different sizes. For each of these partitions, we analyse the relationship between the edge weights \textit{within} co-agglomeration communities and the corresponding edge weights in the customer-supplier and labour networks. We uncover \textit{regions} of the co-agglomeration network - clusters of co-located industries - dominated by labour and/or value chain agglomeration channels. This result has implications for network based diversification models, introduced in the next section, which simulate dynamics on a co-agglomeration network in order to describe economic branching processes. 

As a final validating step, we investigate the relationship between the mean number of years of education (averaged across industries within a cluster) and our cluster estimates for reliance on labour and customer-supplier linkages. It is expected that labour-dominated clusters would employ workers with a higher level of education. In agreement with this hypothesis, we find that years of education correlate positively with our estimates for the labour-sharing channel, and correlate negatively with our estimates for the IO channel.

%%%%%%%%%%%%%%%%%%%%%%%%%%%%%%%%%%%%%%%%%%%%%%%%%%%%%%%%%%%%%%%%%%%%%%%%%%%%%%%%%%%%%%%%%%%
\subsection{Industrial growth and diversification models \label{section_div}}

An emerging perspective sees diversification as a path dependent process, whereby the growth and appearance of new economic activities (industries) in a place is dependant on the local availability of relevant capabilities \citep{nelson1982, frenken2007, hidalgo2007}. These capabilities include, but are not constrained to, skilled labour, physical infrastructure and other necessary inputs for production. 

This perspective has emerged from the (recently converging) fields of economic complexity and evolutionary economic geography, and particularly focuses on the role of local capabilities (often in the form of worker know-how) in growth and diversification process \citep{nelson1982, frenken2007, hidalgo2007}. Hence, quantifying local capabilities or know-how (termed `economic complexity’ by \citet{hidalgo2009}), and identifying new activities that are `proximate’ in terms of their capability requirements \citep{frenken2007, hidalgo2007}, is key to the development of diversification and development strategies and models. 

Within this framework, cities move into new economic activities that share existing capabilities in a path dependent manner. This process can be modelled using an industry network, where edges represent capability-overlap \citep{hidalgo2007, neffke2011regions}. This network can be seen as an economic `map’ or `landscape’: the position of a city (i.e., the subgraph of its industries) constrains its future development path. More specifically, the diversification of a city into new industries can be modelled by a diffusion process on the industry network with initial condition governed by the initial set of industries in the city \citep{hidalgo2007, neffke2011regions, hausmann2014, formality}. Hence, while centrally located cities share capabilities with many potential new industries, peripheral cities have fewer options.

This network approach to modelling diversification processes was pioneered by \citet{hidalgo2007}, and focused on describing the diversification paths of countries as they develop and move into new, more sophisticated products. The underlying network of products (the \textit{Product Space}) is a product-country analogue of the (industry-city) EG network introduced in the previous section. In other words, in the diversification literature, the EG network has been proposed as a general measure of capability overlap or `industry relatedness' \citep{hidalgo2007, hausmann2014}, which does not distinguish between different types of linkage, and forms the `landscape' on which to model growth and diversification dynamics. Hence, our approach to uncovering the relationship between co-agglomeration patterns and the channels of labour sharing and customer/supplier linkages at an industry cluster scale has clear implications for this class of diversification models. We expect that diversification dynamics will be better captured by skill linkages in some regions of the network (as uncovered by our analysis), and better modelled by IO linkages in other regions of the network. DNO provide evidence that employment growth dynamics at the city-industry level can be improved by incorporating industry-level heterogeneous linkages, but there remains future work with respect to our cluster estimates.

Modelling diversification paths using industry networks is well-established in the `related diversification' literature, and has been deployed to model growth and diversification processes across a range of spatial scales \citep{frenken2007, neffke2011regions, Neffke2013SkillRelatedness}, and study a wide range of questions around local growth paths, including employment and formality rates, and firm and sector entry (\citealp{hausmann2014, formality}).

\section{Data and Metrics}

Here we will briefly review the technical definitions of the industry similarity metrics used by EGK and DNO as introduced in Section \ref{sect_sim}, and provide information on data sources. For the new analysis presented in Section 4, we will use the same data as DNO. 

\textit{Co-agglomeration}

As proposed by \citet{ellison1994geographic}, we define a co-agglomeration index for industries $i$ and $j$:
\begin{equation}
\label{eqnEG}
EG_{ij}=\frac{\sum_{r=1}^{R}(s_{ir}-x_r)(s_{jr}-x_r)}{1-\sum_{r=1}^{R} x_r^2},
\end{equation}
where $s_{ir}$ is the employment share of industry $i$ in region $r$, and $x_r$ is the mean share of employment in region $r$. As mentioned by DNO, this index has the advantage of not being affected by the size distributions of firms in various industries nor by the level of spatial aggregation. 

For the analysis undertaken by EGK, 1987 employment data from the US Census Bureau’s Census of Manufacturing (122 three-digit SIC manufacturing industries) was used. In the case of DNO, 2003 employment data from the US Bureau of Labour Statistics (283 four-digit NAICS industries including both manufacturing and services, and 939 metropolitan areas) is used. 

\textit{Labour pooling}

Next, we measure industry similarity with respect to labour pooling using occupational data as proposed by \citet{farjoun1994beyond}. If $E_i$ is a vector of employment by occupation for industry $i$, we compute the Pearson correlation between $E_i$ and $E_j$:
\begin{equation}
\label{eqnL}
L_{ij}=corr(E_i,E_j).
\end{equation}
Observe that this metric is symmetric with respect to $i$ and $j$ and does not depend on the sizes of the respective industries. 

For the analysis undertaken by EGK, 1987 industry-occupation employment data was taken from the National Industrial-Occupation Employment Matrix (NIOEM) published by the US Bureau of Labor Statistics (277 occupations). In the case of DNO, 2002 industry-occupation employment data was taken from the US Occupational Employment Statistics (734 occupations). 

\textit{Customer-supplier linkages}

In order to study customer-supplier relationships between industry pairs, we use an Input-Output matrix with entries $i,j$ corresponding to the value of goods and services that industry $j$ sources from industry $i$. Normalising in both directions to account for the relative importance of industries as a buyer and seller, we create a symmetric $IO$ proximity matrix:
\begin{equation}
\label{eqnIO}
IO_{ij}=\max \Big(\frac{IO_{ij}}{\sum_k IO_{kj}}, \frac{IO_{ji}}{\sum_k IO_{kj}},\frac{IO_{ij}}{\sum_k IO_{ik}},\frac{IO_{ji}}{\sum_k IO_{ik}} \Big).
\end{equation} 

For the analysis undertaken by EGK, the 1987 Benchmark Input-Output Accounts published by the Bureau of Economic Analysis (BEA) was used. In the case of DNO, analogous tables provided by the BEA for the year 2002 are used.

\textit{Knowledge spillovers}

We can estimate inter-industry knowledge similarity using cross-industry patent citations - specifically, the extent to which technologies associated with industry $i$ cite technologies associated with industry $j$, and vice versa. Using a similar approach to Eq.~\ref{eqnIO}, we compute: 
\begin{equation}
\label{eqnIO}
K_{ij}=\max \Big(\frac{X_{ij}}{\sum_k X_{kj}}, \frac{X_{ji}}{\sum_k X_{kj}},\frac{X_{ij}}{\sum_k X_{ik}},\frac{X_{ji}}{\sum_k X_{ik}} \Big).
\end{equation} 
where $X_{ij}$ is the number of citations from patents associated with
industry $i$ to patents associated with industry $j$. 

Both EGK and DNO make use of the NBER patent citations dataset \citep{hall2001nber} to compute this matrix. 

\textit{Education}

Finally, in Section \ref{sect_edu} we use US micro-data (2002) from the Integrated Public Use Microdata Series (IPUMS), the world's largest individual-level population database, to compute the mean number of years of education for workers in US industries.

\section{Unravelling the forces of agglomeration}

In this section, we explore the methodological approach and results of EGK and DNO. In summary, EGK show that input-output linkages are the dominant channel for manufacturing industries using data from 1987. Using more recent data, DNO show that labour sharing has become over time an increasingly important channel, and the dominant channel in service industries. Additionally, while EGK view these channels as homogeneous across all industries, DNO find significant heterogeneity between industries.

\begin{table}[t!]
    \centering
    \caption{OLS univariate regressions from EGK and DNO (Eq.~\ref{betaseqn1})}
    \begin{tabular}{c c c c c c c c}
    \hline \hline
         & EGK (manu) & & DNO (manu) & & DNO (serv) & & DNO (all)\\
         \hline
         L& 0.106 &  &0.164 &&  0.275&& 0.175 \\
         & (0.016) &  &(0.018) &&  (0.037)&&  (0.007)\\
         N& 7381 &  &3655 &&  5360 &&  16836\\
         $R^2$& 0.011 & & 0.035 && 0.018 &&  0.034\\
         \hline
         IO&0.167 & & 0.177 &&  0.171&&  0.138\\
         & (0.028) & & (0.032) &&  (0.028)&&  (0.015)\\
         N& 7381 & & 3655 &&  5360&&  16836\\
         $R^2$&0.028 & & 0.060 &&  0.016&& 0.022\\
         \hline
         K&0.100 & & 0.109  &&  --&&  --\\
         & (0.016) & & (0.013)  & &-- & & --\\
         N& 7381 & & 3655 & &-- && -- \\
         $R^2$&0.010 & & 0.022 && -- & &--\\
         \hline
    \end{tabular}
    \label{tab1}
\end{table}

\subsection{Homogeneous forces (EGK) \label{sect_EGK}}

Using manufacturing data from the US (1987), EGK sought to study the relationship between industry co-location, and shared labour, customer-supplier and knowledge requirements via estimation of the coefficients $\beta_{Z}$ of a simple linear model: 
\begin{equation}
EG_{ij}= \alpha + \beta_{Z} Z_{ij} +\epsilon_{ij},
\label{betaseqn1}
\end{equation}
across all pairs $i,j$ for channels $Z\in\{L,IO,K\}$. Table~\ref{tab1} shows a much stronger ($>50\%$) relationship between co-agglomeration patterns and value chain linkages as compared to labour or knowledge linkages. 

Following the approach of EGK, DNO investigated the contribution of labour pooling and customer-supplier relationships to co-location using more recent data from the US (2002). This study does not include knowledge spillovers, and natural resource availability, as was incorporated into previous work. DNO showed, using the same estimation model, that labour sharing is increasingly an equal driver of agglomeration in manufacturing industries, and a more significant driver of agglomeration in services (see Table~\ref{tab1}). 

Applying this approach to time-series data (1910-2010), DNO showed that the influence of value chain linkages has consistently decreased over time since before the 1940s. Furthermore, their analysis showed that while the importance of labour to firm co-location has declined somewhat in recent years, it is still significantly higher than it was in 1950. If previous trends are to persist, this study suggests that value chain linkages will continue to lose importance, while labour pooling will remain a driving factor of co-agglomeration.

\begin{figure}[t!]
    \centering
    \includegraphics[width=.9\linewidth]{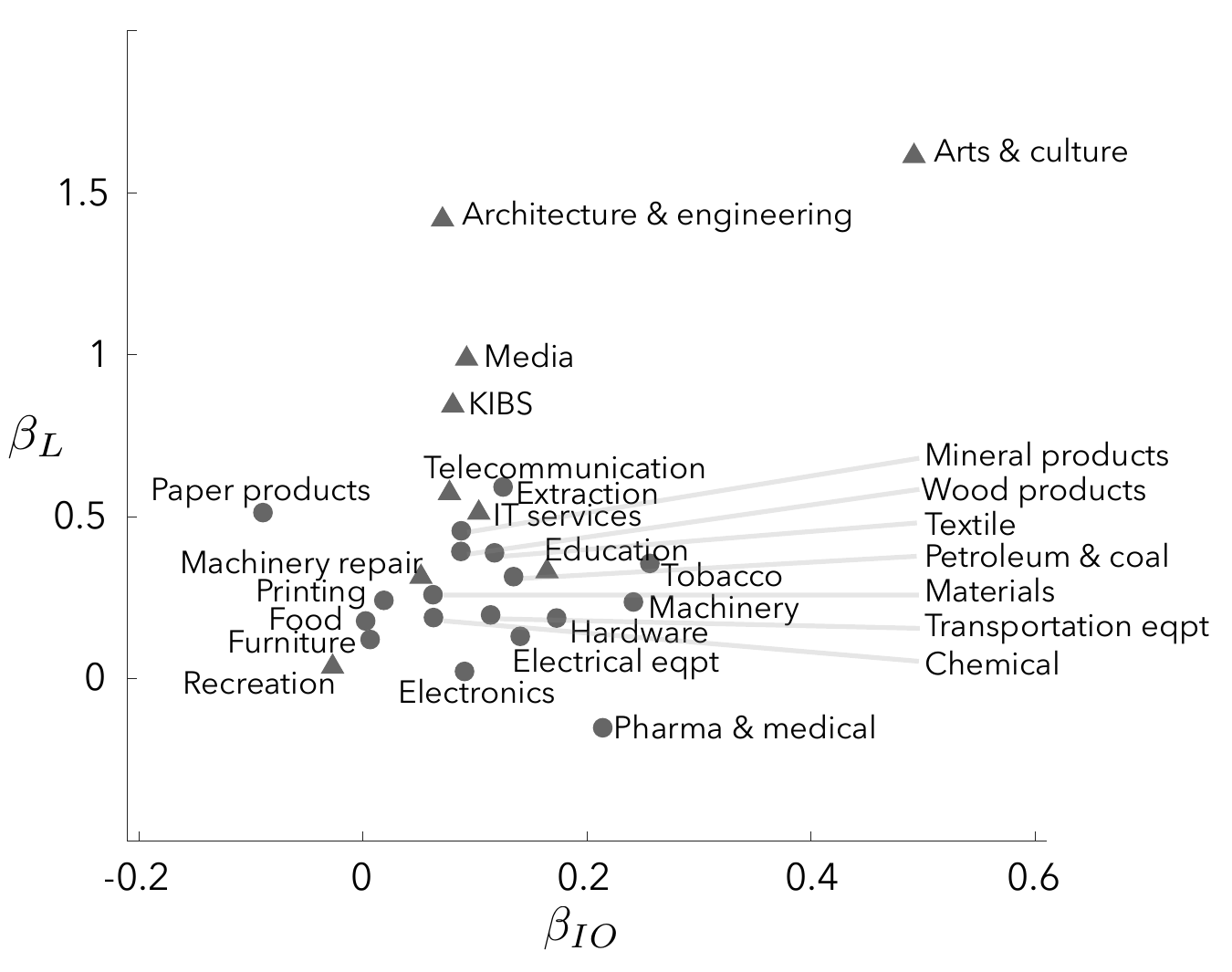}
    \caption{Average estimates of labor pooling effects ($\beta_L^i$) and value chain effects ($\beta_{IO}^i$) from Eq.~\ref{betaseqn2} are displayed for 27 sectors, demonstrating wide variation in the importance of each of these Marshallian channels across different types of industries. Estimates for services are shown using triangles, while those for manufacturing industries are shown using circles. We observe that service industries tend to be more labour pooling driven than value chain driven. \label{fig_betas}}
\end{figure}

\subsection{Heterogeneous forces (DNO) \label{sect_DNO}}

As discussed above, it is intuitive that the relative importance of each Marshallian channel will be different for each industry. DNO explore this idea, allowing the relationship between co-location and shared labour as well as value chain linkages to vary by industry. Specifically, DNO estimate the coefficients $\beta_{Z}^i$ for each industry $i$ via univariate regressions of the form:
\begin{equation}
\label{betaseqn2}
EG_{ij}= \alpha_i+\beta_{Z}^iZ_{ij} + \epsilon_{ij}
\end{equation}
across all industries $j$ for channels \emph{Z}$\in\{L,IO\}$. Essentially, these coefficients capture the relationship between co-location and labour/value chain similarities for linkages between an industry $i$ and all other industries $j$. 

Using this approach, it is possible to estimate labour and value chain coefficients for \textit{each} industry, enabling us to assess the relative contribution of these channels for individual industries. We plot the estimated coefficient pairs $\beta_{IO}^i$ and $\beta_{L}^i$ (averaged for 27 sectors) in Figure~\ref{fig_betas}. This figure is a reproduction - with permission - of Figure 3 in \citet{diodato2018industries}. 

We observe a distribution of coefficients such that knowledge-intensive industries such as design and accounting rely on specific labour availability, whereas industries that rely on physical inputs such as agriculture, construction and manufacturing depend more on value chain sharing (Fig.~\ref{fig_betas}). Moreover, we observe more generally that service industries tend to have higher labour sharing dependence than value chain dependence ($\beta_L > \beta_{IO})$, whereas manufacturing industries have quite varying ratios of $\beta_L$ to $\beta_{IO}$.

Hence, by calculating industry-specific Marshallian agglomeration forces, DNO showed that labour-sharing is the most important motive behind contemporary location choices of services, although value chain linkages still explain much of the co-location patterns in manufacturing. 

%%%%%%%%%%%%%%%%%%%%%%%%%%%%%%%%%%%%%%%%%%%%%%%%%%%%%%%%%%%%%%%%%%%%%%%%%%%%%%%%%%%%%%%%%%%
\section{Industry clusters exhibit heterogeneous drivers of agglomeration patterns}

In this section, we explore the idea that industry \emph{clusters} - that is groups of industries co-located across many cities - represent a natural intermediate scale at which to examine heterogeneous drivers of agglomeration patterns. Specifically, we identify industry clusters in the co-agglomeration \emph{network} which exhibit heterogeneous dependence on each of the Marshallian channels, captured via labour and customer-supplier linkages as before.

\begin{figure}[t!]
    \centering
    \includegraphics[width=.9\linewidth]{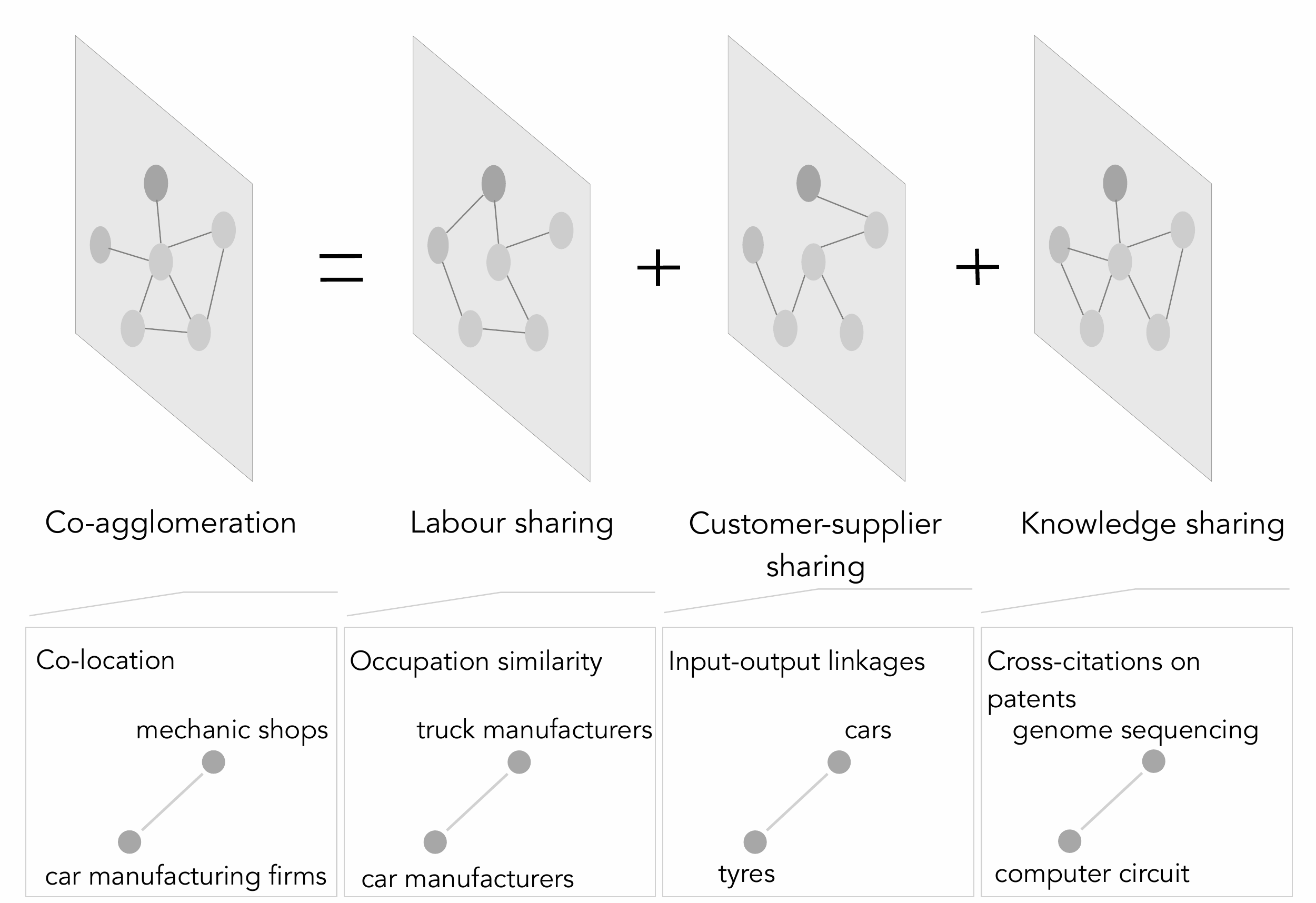}
    \caption{Pairwise industrial co-agglomeration patterns can be represented as a network, with edges estimated via the co-location patterns of industries in cities. Schematically, co-agglomeration patterns are a composite of labour, customer-supplier and knowledge sharing patterns - measured via occupation similarity, input-output linkages and patent citations respectively.}
    \label{fig:schematic}
\end{figure}

%%%%%%%%%%%%%%%%%%%%%%%%%%%%%%%%%%%%%%%%%%%%%%%%%%%%%%%%%%%%%%%%%%%%%%%%%%%%%%%%%%%%%%%%%%%
\subsection{The Markov Stability framework for community detection}

The general task of community detection is to find a node partition (i.e., a split of the nodes into communities) such that nodes in the same community have more edges than would be expected in a random graph with the same degree distribution. Here, we use the Markov Stability algorithm \citep{delvenne2010stability} to perform this task on the EG network. Below we briefly review the Markov Stability framework, and refer a reader to the original paper for more details.

In general, we can describe a (discrete) random walker diffusion process on a graph under the updating rule: 
\begin{equation}
p_{t+1}=p_t M
\end{equation}
where $p_t$ is the (normalized) probability vector ($p_t(i)$ is the probability a random walker is on node $i$ at time $t$), $M=D^{-1}EG$ is the transition matrix ($D$ is a zero matrix with the vector of node degrees of EG - denoted deg$(EG)$ - on the diagonal). The stationary distribution of this process is given by $\pi=$deg$(EG)/$sum(deg($EG$)). 

We want to quantify the clustering of these dynamics for nodes within communities. We can encode a node partition in a matrix $H$ such that $H(i,c)=1$ if node $i$ is in community $c$ (and otherwise $H(i,c)=0$). The clustered autocovariance matrix of the diffusion process is defined as:
\begin{equation}
\label{defR}
R(t)=H^T [\Pi M^t- \pi^T \pi] H,
\end{equation}
where $\Pi = $ diag $(\pi)$. Observe that $(\Pi M^t)_{uv}$ represents the probability that a random walker who started in community $u$ ends up in community $v$ at time $t$, and $(\pi^T \pi)_{uv}$ is the probability that two independent random walkers are in $u$ and $v$ at stationarity. The diagonal entries of $R(t)$ therefore represent the probability that a random walker remains in their initial community after $t$ time-steps, and hence the \textit{stability} of a partition at time $t$ is defined as:
\begin{equation}
\label{defr}
r(t)=Trace(R(t)).
\end{equation}
If a network's community structure is well-defined, then a random walker is highly likely to remain in the community in which it started - therefore, we seek a partition matrix $\hat{H}$ that maximises $r(t,\hat{H})$ on the set of all the possible partitions. This problem is NP-hard, and we therefore we use the heuristic Louvain method to solve it. 

Louvain's method \citep{blondel2008fast} works as follows. The method first assigns each node to its own community. Then, for each node, it considers merging it with each of its neighbours (e.g., merging their communities) one by one. Merging occurs if this results in an increased value of the optimization criteria, in this case stability as given by Eq.~\ref{defr}. The process is repeated until no increase in stability can be achieved. 

This algorithm is stochastic in the sense that it does not have a unique solution, and will not necessarily return the same partition on each run. Hence, for each Markov time $t$, we run Louvain's algorithm 1000 times. We then compare each pair of partitions found using the variation of information \citep{meilua2007comparing}, and take the average across all pairs. If the mean variation of information is low at some $t$, the obtained partitions are similar, signalling robustness.

\begin{figure}[t!]
    \centering
    \includegraphics[width=1.1\linewidth]{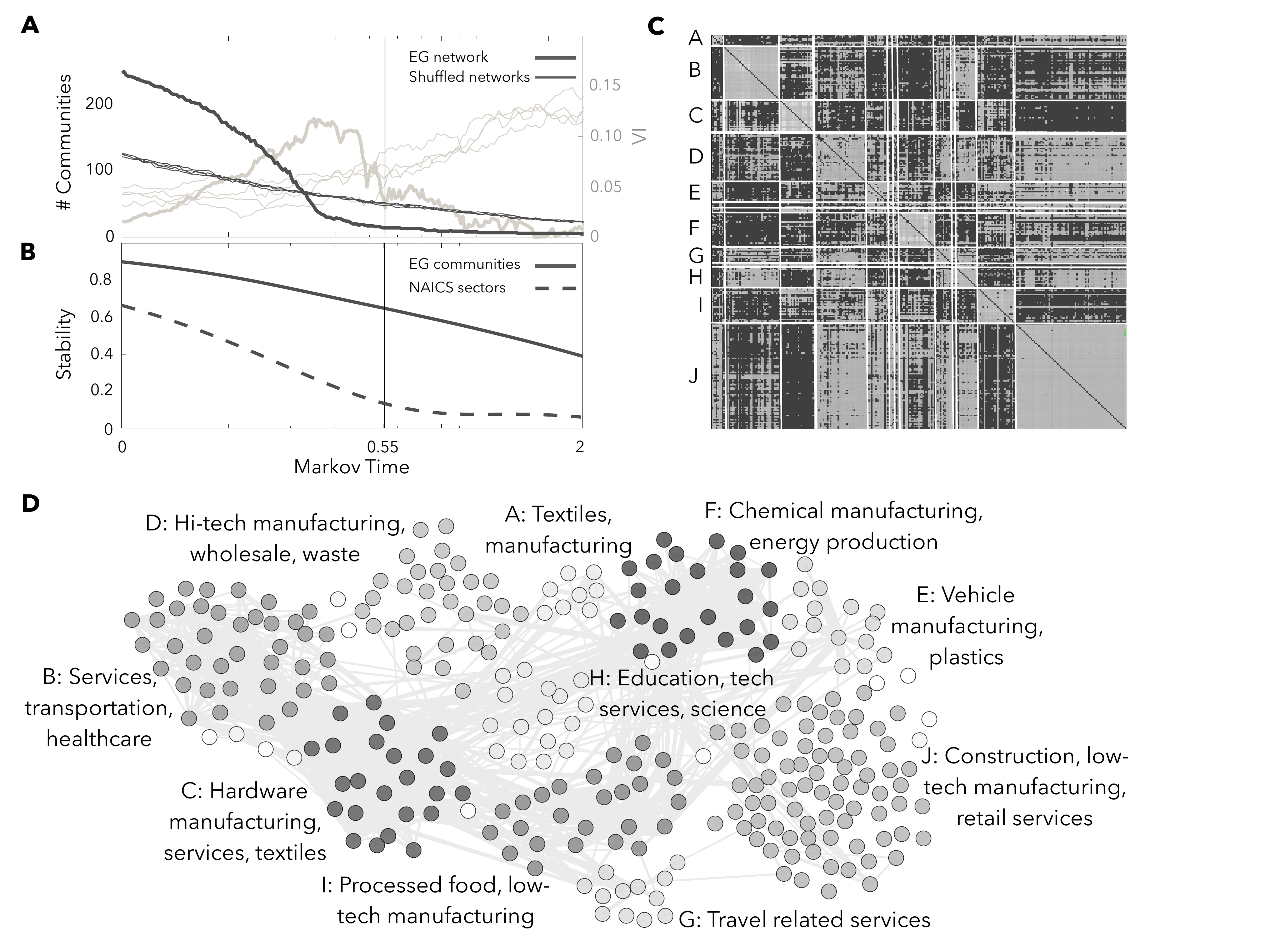}
    \caption{[A] As the Markov time increases, the number of communities decreases (communities become larger). The Variation of Information (VI), a measure of partition robustness, exhibits a number of local minima. We compare these metrics for the EG network, and 4 shuffled networks (the EG network with edges randomly shuffled). The vertical line corresponds to the partition $P_{14}$ visualised in subfigures C and D. 
    [B] Stability, a measure of partition quality, is shown for the EG network and a partition corresponding to 23 official NAICs 2-digit sectors. The stability method identifies more robust communities compared to the official industry grouping.
    [C] Heatmap of the $EG$ matrix, organized according to the partition $P_{14}$. Paler entries correspond to higher edge weights. Community structure is apparent via the relative density of within-community edges (diagonal blocks). 
    [D] Visualisation of the $EG$ network, with nodes coloured according to the partition $P_{14}$, and labels assigned from manual inspection of industries. The top $2\%$ of edges are displayed.}
    \label{fig3}
\end{figure}
%%%%%%%%%%%%%%%%%%%%%%%%%%%%%%%%%%%%%%%%%%%%%%%%%%%%%%%%%%%%%%%%%%%%%%%%%%%%%%%%%%%
\subsection{Community structure in the EG network \label{sec:inm}}

We apply the Markov stability framework for community detection on the EG network. Communities in the EG network correspond to groups of industries that tend to co-locate across a large number of cities. 

We find (Fig.~\ref{fig3} A) multiple Markov times at which the computed partition is robust (e.g. times corresponding to local minima in the variation of information, \emph{VI}). These correspond to node partitions on a range of scales, from many small communities to few large communities. In order to explore the network's community structure, we examine in more detail the partition corresponding to the local $VI$ minimum at $t=0.55$. This partition, which we call $P_{14}$, contains 14 distinct communities, of which 10 have $\ge5$ nodes (Fig.~\ref{fig3} C and D). At this Markov time, many communities have intermediate size, with 8 having between $15$ and $32$ industries, while one community is quite large ($75$ industries). 

Community $B$ contains 38 industries, most of which are different types of services - in particular healthcare, education, and finance. Because of the tendency of service industries to require specialised labourers, we might expect that the observed co-location patterns are driven by labour linkages. On the other hand, community $E$ contains seven industries involved in plastics manufacturing or machinery, and four industries related to motor vehicle manufacturing/repair. In this case, we might hypothesise that customer supplier relationships drive co-location patterns, as manufacturing/repair of motor vehicles may co-locate with plastic/metal manufacturing firms that supply necessary components. In the next section we will investigate these hypotheses.  

\begin{figure}[t!]
    \centering
    \includegraphics[width=0.8\linewidth]{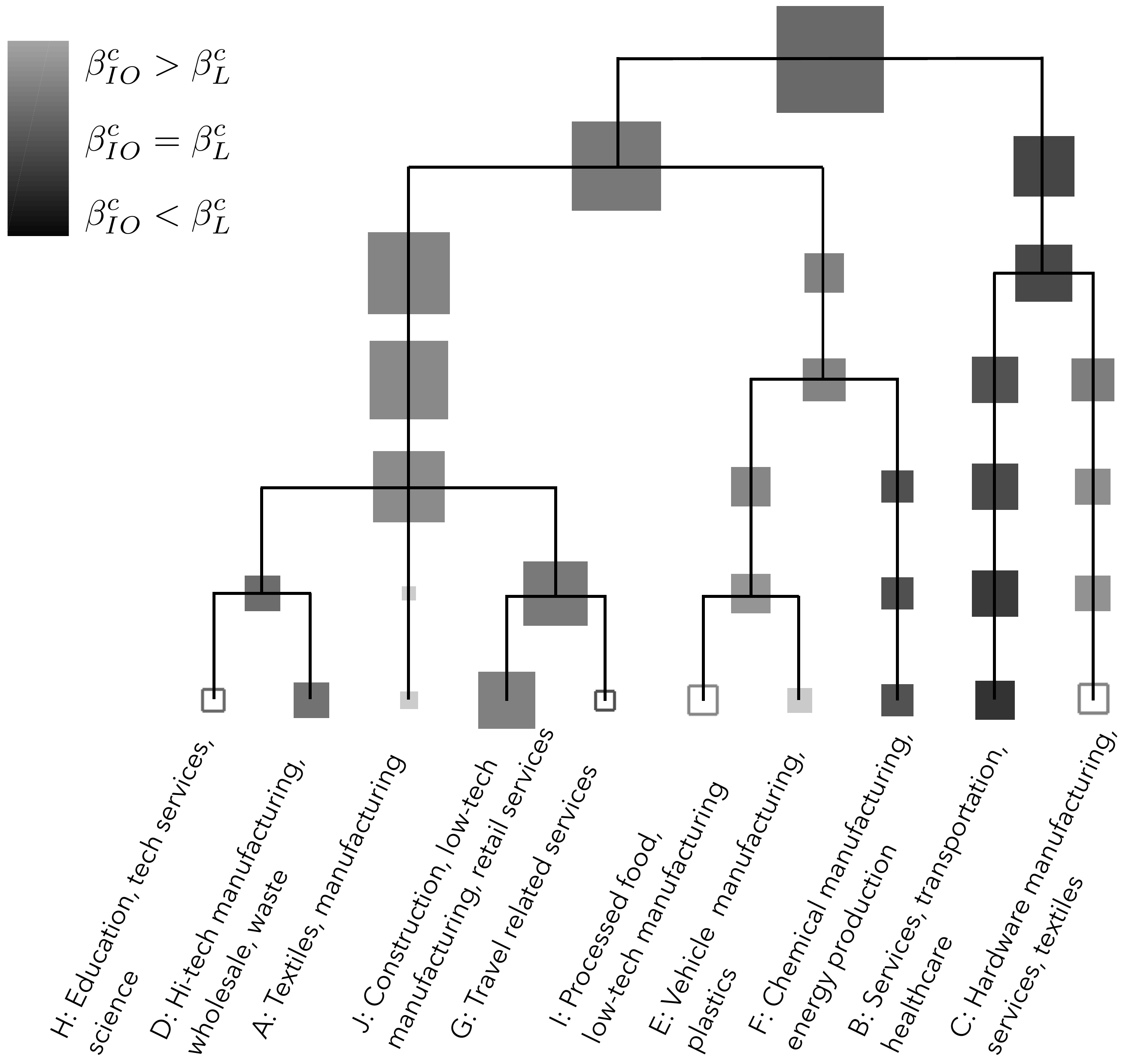}
    \caption{Dendrogram of hierarchical community structure for seven partitions of the EG network, starting from the whole network in one community (root) to partition $P_14$ of size 10 communities (leaves). The size of the square corresponds to community size, and colour corresponds to dependence on the labour and/or IO agglomeration channel (as shown in legend).
    We observe at the coarsest division one community dominated by labour sharing and another dominated by value chain linkages. As the partitions become finer, we see a wide distribution in the importance of the two channels in different communities (open boxes represent communities in which neither $\beta_{IO}^c$ nor $\beta_{L}^c$ are significant).}
    \label{fig4}
\end{figure}

In order to visualise the multi-scale community structure of the EG network, we choose seven representative partitions (for Markov times greater than $t=0.55$) to construct a dendrogram (Fig.~\ref{fig4}). These range from the whole network (top) to the ten community partition (leaves) discussed above ($P_{14}$). Because our partitions aren't strictly hierarchical, we employ a simple majority voting scheme to assign communities to a 'parent' community in a coarser partition.

In order to further demonstrate the robustness of the communities we detect in the EG network, we compare these to communities detected in random networks derived from shuffling the edges in the EG network. Fig.~\ref{fig3} A shows that for sufficient Markov time, the $VI$ for the EG is significantly lower than for the shuffled networks. Note that in subsequent analysis, we only consider partitions for $t\ge 0.55$.

Finally, we ask whether our communities correspond to a better partition of the EG network than administrative ``communities'' given by two-digit NAICS subsector classification. This latter comparison is particularly important as it validates to some extent the need for community detection to identify 'exogenous' industry clusters. We see in Fig.~\ref{fig3} B that the stability (quality) of the detected EG communities greatly exceeds that of the NAICS sector partitions, especially for $t\ge 0.55.$

%%%%%%%%%%%%%%%%%%%%%%%%%%%%%%%%%%%%%%%%%%%%%%%%%%%%%%%%%%%%%%%%%%%%%%%%%%%%%%%%%%%
\subsection{Industry clusters display heterogeneous preferences for labour and value chain agglomeration channels}

Previous analysis has been limited to measuring the relative importance of Marshallian channels either at the individual industry level or at the whole network level. Here, we argue for a meso-scale approach, with the intuition that dependence on each of the Marshallian channels varies not by individual industries but by clusters of co-located industries, which can also be conceptualized as regions in a capability 'landscape' (see Section 1.5).

Specifically, we seek to quantify the dependence of each community on the labour and value chain agglomeration channels. That is, in place of Eq.~\ref{betaseqn2}, we let the respective channel coefficients vary by community $c$ (instead of industry) via the regression:
\begin{equation}
\label{betaseqn3}
{EG}_{ij}= \alpha_{c}+\beta_Z^c Z_{ij} + \epsilon_{ij},
\end{equation}
for industry pairs $i,j \in c$, and channels $Z\in\{IO,L\}$. Essentially, the coefficients $\beta_{L}^c$ and $\beta_{IO}^c$ capture the relationship between co-location and labour similarities/value chain linkages between industry pairs $i$ and $j$ within community $c$. 

We estimate the coefficients in the OLS regression (Eq. \ref{betaseqn3}) for various Markov times. We can visualize the progression from homogeneity to heterogeneity in the channel coefficients via the dendrogram in Fig.~\ref{fig4}. From this analysis, we see (from top to bottom) a shift from the simple labour vs. input output community partition to a more complex structure composed of heterogeneous network regions. At the first split, the network decomposes into two communities, one driven by input/output linkages (left) and the other driven by labour sharing (right). The input/output branch then splits into two sub-branches which are also input/output dominated, but a series of splits reveals heterogeneous channel importance at high resolution. For instance, we see that community $E$ is dominated by input/output linkages (in line with our expectations), whereas community $F$ (sophisticated chemical manufacturing and energy production) is dominated by labour sharing. The labour branch (top right) also splits into two sub-branches, one dominated by input/output and the other dominated by labour sharing. This latter community (driven by labour linkages), is community $B$, which is composed of service industries as explored earlier.

\begin{figure}[t!]
    \centering
    \includegraphics[width=\linewidth]{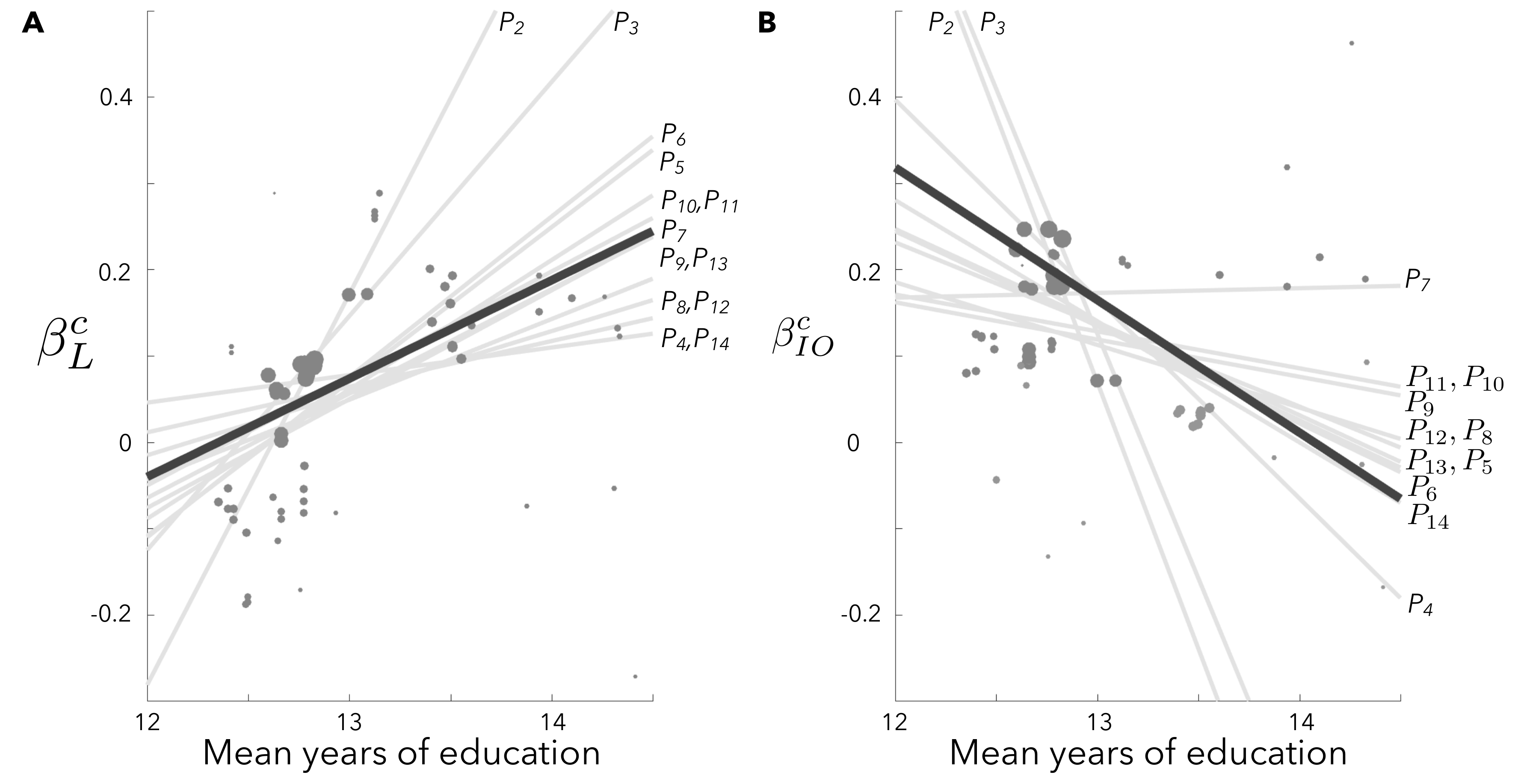}
    \caption{Estimates of $\beta_L^c$ and $\beta_{IO}^c$ vs mean years of education for 11 different community partitions (points sized by community size). Pale lines correspond to weighted least squares fits for each partition (see Eq.~\ref{eq:education}), while the dark line corresponds to a weighted fit across all partitions (column three of Table 2). We observe a clear tendency for communities that are more biased towards the labour sharing agglomeration channel to have more educated workers. Communities that are more biased towards the value chain sharing agglomeration channel tend to have less educated workers.}
    \label{fig5}
\end{figure}

%%%%%%%%%%%%%%%%%%%%%%%%%%%%%%%%%%%%%%%%%%%%%%%%%%%%%%%%%%%%%%%%%%%%%%%%%%%%%
\subsection{Preference for labour pooling correlates with education level of workers \label{sect_edu}}

Here, we seek to investigate the relationship between the strength of labour pooling in co-agglomeration patterns and years of worker education at the cluster level. We expect that clusters dominated by service industries which tend to co-locate based on shared skill requirements will also tend to employ more highly educated workers.

Employing data from IPUMs, we calculate for each industry the average number of years of education for workers - and then for each community the average number of years across industries (denoted $ed_c$). Then, we examine the relationship between $ed_c$ and the community-level coefficients $\beta_{IO}^c$, $\beta_L^c$. Specifically, we perform regressions of the form,
\begin{equation}
\label{eq:education}
    \beta_{Z}^c=\alpha+ b_Z ed_c+\epsilon_{c}, 
\end{equation}
for all communities $c$ in some partition and $Z\in\{L,IO\}.$    

We perform this regression for many different partitions, corresponding to a range of Markov times. Specifically, we denote by $P_k$ the partition corresponding to the earliest Markov time at which $k$ communities are detected, and (independently) perform regression (\ref{eq:education}) for partitions $P_2,P_3,...,P_{14}$. Rather than performing an ordinary least squares regression (or slope calculation in the $P_2$ case), we take a weighted least squares approach to account for community size (i.e., the diagonal entries of the errors covariance matrix are given by the relative sizes of the communities).

In Fig.~\ref{fig5}, we plot fits of the weighted regression for $Z=L$ (left) and $Z=IO$ (right). In isolation, each of the fits (pale lines) are not terribly meaningful (especially for high Markov time given the small number of communities). However, we note that for all partitions the computed values of $b_{L}$ are positive, and all computed values of $b_{IO}$ are negative except in one case ($P_7$). 

In order to pool the results of these partition-level fits, we perform a second set of regressions that considers all communities \emph{across all partitions}. We first perform an \emph{OLS} regression (column 1 of Table \ref{tab2}) - in which all communities are weighted equally - and find a significant positive estimate for $b_L$ as well as a significant negative estimate for $b_{IO}$ ($p<.01$ for both). That is, we see a strong relationship between the strength of labour pooling in co-agglomeration patterns and years of worker education, as expected, and as well a strong negative relationship between the strength of customer/supplier relationships in co-agglomeration patterns and years of education. Moreover, we find a particularly good linear fit ($R^2>0.4$) for the relationship between $\beta_L^c-\beta_{IO}^c$ and education level. Hence, industry communities that tend to have a very strong dependence on labour pooling relative to input/output linkages on average require more years of education.

We also perform a weighted version (\emph{WLSI}) of this regression that equalizes the total error weights of each partition, and a second weighted version (\emph{WLSII}) in which error weights are proportional to community size (partitions again have equal total weight), again finding significance at the $1\%$ level in both cases (columns 2-3 of Table~\ref{tab2}). This latter pooled fit is shown via a dark line in Fig.~\ref{fig5} A-B.

Some of our estimates for $\beta_L^c$ and $\beta_{IO}^c$ were statistically insignificant (i.e., we cannot reject the null hypothesis that these coefficients are zero). These insignificant estimates occur mainly in small communities in the finer partitions of the network ($P_{14}$), as seen in Fig.~\ref{fig3}. In order to address this issue, we perform versions of these three regressions (OLS, WLSI, WLSII) for the cases in which statistically insignificant values of $\beta_{IO}^c$ and $\beta_L^c$ are set to zero. For columns 4-6 (7-9) in Table~\ref{tab2}, we set the coefficients that were statistically insignificant at the $10\%$ ($5\%$) level to zero. We find significant and comparable results across all cases. 

    \begin{table}[t!]\centering
    \caption{Relationship between community-level L/IO estimates and mean years of education.}
    \scriptsize{\begin{tabular}{c c c c c c c c c c}
    \hline
        \hline
        Dependent variable: $\beta_L$ 
         & \textbf{OLS} & \textbf{WLSI} & \textbf{WLSII} & \textbf{OLS$\dagger$} &\textbf{WLSI$\dagger$} &\textbf{WLSII$\dagger$} & \textbf{OLS$\dagger\dagger$} &\textbf{WLSI$\dagger\dagger$}&\textbf{WLSII$\dagger\dagger$} \\ \hline
        Mean yrs education         &         0.0715   & 0.0794 &  0.0959& 0.0770&    0.0780&    0.1130  &  0.0705&    0.0712 &   0.1037\\
                    &            (0.0218)&(0.0221)&(0.0185)&(0.0184)&(0.0193)&(0.0197)&(0.0174)&(0.0183)&(0.0185) \\
         N           &       80& 80  &80&  80 & 80 & 80&  80& 80  & 80 \\          
        R$^{2}$           &               0.1212   & 0.1420   & 0.2553 &   0.1837    &0.1728  &  0.2960  & 0.1743   & 0.1629    &0.2867 \\
        \hline 
        \\
         \hline
        \hline
        Dependent variable: $\beta_{IO}$ 
         & \textbf{OLS} & \textbf{WLSI} & \textbf{WLSII} & \textbf{OLS$\dagger$} &\textbf{WLSI$\dagger$} &\textbf{WLSII$\dagger$} & \textbf{OLS$\dagger\dagger$} &\textbf{WLSI$\dagger\dagger$}&\textbf{WLSII$\dagger\dagger$} \\ \hline

        \hline  
       Mean yrs education   &          -0.1657&   -0.1536&   -0.1795&   -0.1675&   -0.1580&   -0.0959&   -0.1659&   -0.1567&   -0.0939 \\
                    &             (0.0304)&(0.0298)&(0.0304)&(0.0297)&(0.0294)&(0.0259))&(0.0301)&(0.0297)&(0.0263) \\
         N           &        80&   80  & 80 &  80&   80&   80 &  80&   80  & 80 \\          
        R$^{2}$           &       0.2757   & 0.2538 &   0.3090&    0.2896   & 0.2707   & 0.1491 &   0.2799   & 0.2630 &   0.1401\\
\hline 
        \\
         \hline
        \hline
    Dependent variable: $\beta_L-\beta_{IO}$ 
    & \textbf{OLS} 
    & \textbf{WLSI} 
    & \textbf{WLSII}
    & \textbf{OLS$\dagger$} &\textbf{WLSI$\dagger$} &\textbf{WLSII$\dagger$} 
    & \textbf{OLS$\dagger\dagger$} &\textbf{WLSI$\dagger\dagger$}
    &\textbf{WLSII$\dagger\dagger$} 
    \\ \hline
    Mean yrs education    &    0.2372&    0.2330&    0.1973&    0.2445   & 0.2360   & 0.2089&    0.2363 &  0.2279 &  0.1976\\
        & (0.0280)&(0.0288)&(0.0244)&(0.0301)&(0.0309)&(0.0264)&(0.0310)&(0.0315)&(0.0268)\\
        N   &      80&   80  & 80 &  80&   80&   80 &  80&   80  & 80\\ 
          R$^{2}$  &   0.4797   & 0.4558 &  0.4558   &  0.4583 & 0.4279  &  0.4451&   0.4277    &0.4015 & 0.4098 \\
        \hline
    \hline
    \end{tabular}}
        
    \label{tab2}
    \end{table}

%%%%%%%%%%%%%%%%%%%%%%%%%%%%%%%%%%%%%%%%%%%%%%%%%%%%%%%%%%%%%%%%%%%%%%%%%%%%%%%%%%%%%%%%%%%
\section{Conclusion}

Efforts to disentangle the relative importance of Marshallian channels to the location decisions of firms and industries in cities are building pace. In this chapter we reviewed two important studies which, in turn, sought to unravel the impact of individal agglomeration channels across all industries \citep{EllisonGlaeserKerr2010}, and for individual industries \citep*{diodato2018industries}. 

Building on these studies, we constructed a hierarchical decomposition of the full set of industries into clusters based on co-agglomeration patterns, and estimated the relative importance of individual agglomeration channels for each cluster. We observe a transition from two clusters - one strongly related to the labour channel and the other to the customer-supplier channel - to a wide distribution of channel impacts at a finer partition. Finally, we find robust evidence that clusters exhibiting strong dependence on the labour channel employ more educated workers. 

Our decomposition of the co-agglomeration network into regions dominated by one or another agglomeration channel has implications for diversification models using similar networks. There is evidence \citep{diodato2018industries} that such estimates can improve the predictive power of these models, and so we expect this to be an future avenue worth pursuing. From a policy perspective, a city seeking to diversify into 'related' industries might use our cluster estimates to decide whether to focus on labour-oriented policies (e.g., building up the local skill base), or facilitating inter-firm transactions (e.g., policies aimed at promoting local value chains). 

Much of this analysis focused on two of the three Marshallian channels, labour sharing and customer-supplier linkages. The third, knowledge spillovers, is more difficult to capture across a wide range of industries as patenting activity (normally used to capture R\&D interaction between firms and industries) is concentrated in few industries. Future work might seek to investigate alternative methods and data sources for capturing knowledge spillovers across a broader range of industries and activities. 

Finally, this analysis has been conducted using relatively recent data from the USA. While much academic study of urban economies focuses on highly developed cities (particularly US cities), little is known about whether these patterns are similar or substantially different for developing cities, where employment is often primarily informal \citep{formality}.

\bibliographystyle{elsarticle-harv}
\bibliography{bib}

\end{document}